\documentclass[12pt]{article}

\title{{\small\hfill IMSc/2007/07/08}\\
\textbf{Evading divergences in quantum field theory}}
\author{
H. S. Sharatchandra$^{a}$\footnote{E-mail: sharat@imsc.res.in} \\\\
$^a$ The Institute of Mathematical Sciences, C.I.T. Campus, Taramani P.O.,\\
Chennai 600113, India}
\date{}
\begin{document}
\maketitle
\begin{abstract}
Explicit solution of a Green function in a non-renormalizable toy model 
demonstrates that Green functions of the interacting theory fall off
much faster than at the tree level at large momenta. This suggests 
a method of calculations in quantum field theory which is free of 
divergences.
\end{abstract}
\newpage
%%%%%%%%%%%%%%%%%%%%%%%%%

Since the time Dirac quantized the Maxwell's equations, infinities
have plagued quatum field theory (QFT).
Over the decades they have evoked all kinds of responses. 
A consequence of infinite number of degrees of freedom;
a deficiency of perturbation theory; a result of taking products of
fields at the same space-time point etc.. Nevertheless a pragmatic view of 
renormalization of charges and masses has had remarkable successes.
First this was with the verification of predictions of QED to 
awesome accuracy. Next it was with the
standard model of electroweak and strong interactions, where 
the search for the theory itself was motivated by renormalization.  
The  renormalization 
group analysis of Wilson and others  relate the divergences
to infinite number of scales in the theory  and make connection
with other areas of physics. This approach make the 
divergences inevitable and tackles them in a non-perturbative way.
However it has not been possible to bring quantum gravity into this
fold at present. Many regard this as a fundamental defect of QFT
requiring a new paradigm and a new theory. Planck length is
believed to play a crucial role, either as a fundamental parameter
of a new theory or specifying a drastic change in the nature of space-time,
requiring a new way of handling the theory.
                          
                 In this Note we propose a conventional way out. We propose
that the Green functions  of QFT have to be handled via integral equations. We
illustrate that this approach is free of divergences even in so called
non- renormalization    theories. Tackling QFT through integral equations
is nothing new. The
Schwinger- Dyson equations  define QFT through an infinite set of coupled
non-linear integral equations. They have also been used in many contexts in various
approximations, truncations and for non-perturbative analysis.
We give a different orientation to these equations  here.

               We blame the divergences in QFT on the insistance on
perturbative expansion, in particular the presumption that the Green functions  
of the interacting theory
have the asymptotic behaviour of those of the 
tree approximation (i.e. the semi-classical limit). The very
definition of the Green function  via an integral equation  force an 
asymptotic behaviour  which makes all integrals
converge. We demonstrate this with an explicit solution in a toy model
here.

          Consider elastic scattering of two like particles 'A'
by exchange of field quanta 'B' in the ladder approximation. 
For non-exceptional Euclidean ( and therefore off mass shell)
external momenta, all internal momenta can be safely rotated to be 
Euclidean. We consider  a toy model where the (euclidean) propagators
of 'A' and 'B' quanta are respectively
$ 1/ \sqrt{q}  $
and $ 1/q $ respectively. Here $  \vec q $ is the momentum
in 3-(Euclidean) dimensions  and $ q= | \vec q |$.  With our choice, 
the 1-box diagram  diverges logarithmically,
the 2-box diagram  is linearly  divergent and the N-box
diagram has a superficial degree of divergence (N-1).Thus the model
is like a non- renormalization   theory.

        The sum of the ladder diagrams can be formally specified by
an integral equation . We consider the specific case of total incoming momentum 
being zero. This is not a serious restriction as it doesnot alter 
considerations of ultraviolet 
divergences. We denote the net Green function  excluding the external legs
in this case by $ I(\vec q, \vec q')  $  where
the incoming particles of (Euclidean) momenta $  \vec q  $ and $ - \vec q  $ are
scattered into outgoing particles of momenta $ \vec q' $ and $ - \vec q'  $ respectively.
It satisfies  the integral equation \cite{jms}, 
\begin{eqnarray} 
I(\vec q, \vec q')  =  \frac{1}{|\vec q- \vec q'|}   
+e^2 \int \frac{ d^3 q''}{q"}  \frac{I(\vec q”, \vec q') }
{ |\vec q"- \vec q'|}  
\label{ieq}
\end{eqnarray} 
A perturbation expansion in $e^2$ formally generates the ladder diagrams
with increasing degrees of divergence.
Nevertheless the integral equation, (\ref{ieq}) has a finite and
meaningful solution. The reason is clear. The presumption that
that the full Green function  has the asymptotic behaviour of the 
semi classical Green function  is
wrong. The full Green function  has an asymptotic behaviour such that 
integral in (\ref{ieq}) converges. This behaviour is forced by the
integral equation   itself.

               The integral equation  Eq.\ref{ieq} is precisely that 
for the zero energy Green function of the non-relativistic Rutherford 
scattering problem for two like charges of
charge $e$ and mass 1 unit. There has been an extensive and thorough
investigation of the Green function  for the non-relativistic Coulomb 
problem \cite{cp}.  This explains our   choice of propagators and the toy model.
( We have deliberately chosen a
repulsive interaction as if the quanta 'B' are vector bosons to
avoid tachyons  \cite{jms} and the attendent problem in choosing the boundary
conditions for the Green function.)
                  The integral equation  can be converted into a differential 
equation \cite{jms}
\begin{eqnarray} 
(\nabla_q^2 + \frac{ e^2}{ q}) I(\vec q, \vec q')   =
- 4\pi  \delta^{3}(\vec q- \vec q')
\end{eqnarray} 
A solution of this equation  with an appropriate asymptotic behaviour is the 
solution of  the integral equation Eq.\ref{ieq}.
             It has been realised that the 'momentum space Green function' of the 
Coulomb problem 
\begin{eqnarray} 
G(\vec p, \vec p') = \int d^3 q d^3 q' e^{i \vec q  \cdot \vec p - 
i \vec q'  \cdot \vec p' } I(\vec q, \vec q')  
\label{gpp}
\end{eqnarray} 
has nice properties. (We emphasise that  this terminology of the Coulomb
problem is incorrect 
for our Green function $ I(\vec q, \vec q')$   which is already for the momenta 
$ \vec q, \vec q'$ 
and therefore refer to $G(\vec p, \vec p')$ as MSGF.)
This satisfies the integral equation 
\begin{eqnarray} 
p^2  G(\vec p, \vec p')   + e^2 \int d^3p” \frac{ G(\vec p", \vec p')}
{(\vec p" - \vec p')^2 } =  \delta^{3}  (\vec p  - \vec p' )
\label{iep}
\end{eqnarray} 
Remarkably the perturbation in $e^2$  converges for the MSGF \cite{ad,hol}.

               An explicit solution of the Green function  of the 
Coulomb problem 
has been obtained many times in literature using a variety of
techniques. An elegant way is to use the dynamical symmetry
provided by the Rung-Lenz vector. In the case of the zero energy Green function 
relevant to us, the full symmetry group is $E(3)$, the Euclidean group
in three dimensions\cite{pe}. This suggests the choice of a new variable
\begin{eqnarray}
\vec \xi   =  \frac {e^2 \vec p }{ p^2}
\label{xp}
\end{eqnarray} 
Thus $ \vec \xi $ is obtained from $ \vec p $  by inversion in a sphere of 
radius $  e^2 $. This is a dimensionless variable.  Define 
\begin{eqnarray} 
g(\vec \xi, \vec \xi')   =  \frac{(pp')^4}{ \pi e^6} G(\vec p, \vec p')  
\label{gxx}
\end{eqnarray} 
This is also dimensionless. It satisfies the integral equation 
\begin{eqnarray} 
  g(\vec \xi, \vec \xi')   +  \int \frac{ d^3 \vec \xi" }{  2\pi ^2}
  \frac{  g(\vec \xi, \vec \xi') }{ ( \vec \xi"  - \vec \xi') ^2 }=
 \delta^{3}  ( \vec \xi  - \vec \xi')
 \label{iex}
\end{eqnarray} 

This is easily solved by Fourier transform.
\begin{eqnarray} 
g(\vec \xi, \vec \xi')   = \int_0^\infty  \frac{d^3k}{2\pi} \frac{ k}{ k+1}
e^ {i \vec k  \cdot (\vec \xi  - \vec \xi')}
\label{gx}
\end{eqnarray} 
The leading singularities $g(\vec \xi, \vec \xi')$  as 
$ \vec \xi  \rightarrow \vec \xi' $ can be obtained by writing
\begin{eqnarray} 
\frac{ k}{ k+1 }= 1 -\frac{1}{k } + \frac {1}{k(k+1)}
\end{eqnarray} 
This gives 
\begin{eqnarray} 
g(\vec \xi, \vec \xi')  = \delta^{3} ( \vec \xi i- \vec \xi') 
– \frac {1}{ 2 \pi^2 (\vec \xi - \vec \xi' )^2} 
+  \frac{1}{  2 \pi^2 t} (sin(t) ci(t) + cos(t) si(t)) 
\end{eqnarray} 
where  $t=| \vec \xi  - \vec \xi' |$ and $ci(t), si(t)$ are the cosine and sine
integrals.

The corresponding expansion for
for MSGF  \ref{gpp} is 
\begin{eqnarray} 
G(\vec p, \vec p')   = \frac {1}{ pp'} \delta^{3} ( \vec p  - \vec p') 
– e^2 \frac{1}{p^2 p'^2}  \frac{1}{ ( \vec p  - \vec p')^2} + 
(less singular terms)
\label{pe}
\end{eqnarray} 
This is the expansion of the MSGF to $O(e^2)$. Our interest is 
in $ I(\vec q, \vec q') $, Eq. \ref{ieq}. The first term on RHS of equation 
\ref{pe} gives it a contribution
$ 1/| \vec q  - \vec q'| $ which is the contribution of the tree diagram. The
Fourier transform of the second term on RHS of equation  is log divergent in the
ultraviolet. This matches with the divergence of the 1-box diagram. 
Thus a perturbation theory
fails to be meaningful for  our Green function  $ I(\vec q, \vec q')  $ 
while it gives a convergent series for the Fourier transform 
$ G(\vec p, \vec p') $.

               We now evaluate $I(\vec q, \vec q') $ exactly and demonstarte that we get a
meaningful result. For this we first perform the angular integrations over the 
unit vectors $ \hat p , \hat p' $ using the plane wave expansion,
\begin{eqnarray} 
e^{i \vec p  \cdot \vec q }  = 4\pi \sum i^l j_l (p,q) 
 Y_{lm}^*  ( \hat p) Y_{lm}  ( \hat q)
\end{eqnarray} 
and corresponding expressions 
for $exp(-i \vec p'  \cdot \vec q' ), exp (- i e^2 \vec p  \cdot \vec k
  /k^2)$ and $ exp ( i e^2 \vec p'  \cdot \vec k/k^2)$.Here the sum is over
$l=0,1,2...$ and further for each $l$, over $m= -l, -l+1,...l-1,l$.  
Here $ j_l(x) $ are the spherical Bessel functions related 
by $ j_l (x) =\sqrt {\pi/2x} J_{l+1/2}(x)$  to the usual
Bessel functions.
Performing angular integrations using orthogonality
\begin{eqnarray} 
\int \frac{ d \Omega_{p}  }{ 4 \pi } Y_{lm}^* (\hat p) Y_{l'm'}  (\hat p) 
= \delta _{ll'} \delta_{mm'}
\end{eqnarray} 
gives 
\begin{eqnarray} 
I(\vec q, \vec q')  = \int \frac{ dp }{ p^2} \frac{ dp' }{ p'^2} 
\int  \frac{ dk }{ 2\pi}  \frac{ 1 }{ k+1}
\sum j_l (pq) j_l( \frac{ e^2 k}{p}) j_l(pq) j_l ( \frac{e^2 k'}{p'})
Y_{lm}^* (\hat q) Y_{l'm'}  (\hat q)
\end{eqnarray} 
We may now do the integrations over $p$ and $p'$ variables using \cite{gr}
\begin{eqnarray} 
\int   \frac{ dx}{ x^2 } J_{\nu} (\frac{ a}{x}) J_{\nu} (\frac {x}{b}) 
= \frac{1}{a}  J_{2\nu}  (2\sqrt {\frac{a}{b}}), a,b>0, Re{\nu}> -\frac{1}{2}
\end{eqnarray} 
We get
\begin{eqnarray} 
I(\vec q, \vec q') = \frac{1}{4 \pi e^2}  \int_0^\infty   \frac{k}{ k+1}
 \sum J_{2l+1}  (\sqrt{e^2 kq}) J_{2l+1}  (\sqrt{e^2 k'q'}) 
P_l(  \hat q  \cdot  \hat q')
\end{eqnarray} 
where we have summed over the $m$ variable for each $l$ to get the
Legendre function $ P_l$ of the angle between the vectors
$\vec q, \vec q'$.
In order to carry out the final integration over the variable $k$ we use
\begin{eqnarray}
\int_{0}^{\infty} dx \frac{x}{x^2+c^2} J_{ \nu}(ax) J_{ \nu} (bx) 
= I_{ \nu}(bc)  K_{ \nu}(ac),
0<b<a
\end{eqnarray}
where $ I_{\nu},  K_{\nu} $ are the modified Bessel functions of imaginary
argument.
We get 
\begin{eqnarray}
I(\vec q, \vec q')= \frac{1}{\sqrt{qq'}} \sum_l (2l+1) 
I_{2l+1}(\sqrt {e^2 q_{<}}) K_{2l+1}(\sqrt {e^2 q_{>}}) 
P_l( \hat q  \cdot \hat q')
\label{iqqf}
\end{eqnarray}
where $q_{<}, q_{>}$ are respectively the smaller and larger of the momenta $q,q'$.
This gives the 'non-perturbative' evaluation of the sum of ladder diagrams
each of which (except the tree diagram) is divergent. Using the leading 
asymptotic behaviors for large argument,
\begin{eqnarray}
I_{\nu}(z) \sim \frac{e^z}{\sqrt{2 \pi z}} \\
K_{\nu}(z) \sim \frac{\pi e^{-z}}{\sqrt{2 z}}
\end{eqnarray}
we can read off the behaviour of the  Green function when one of the momenta is
much larger than the other:
\begin{eqnarray}
I(\vec q, \vec q') \sim \frac{e^{-e^2 (q_> - q_<)}}{\sqrt{qq'}}
\end{eqnarray}
We see that the Green function falls off exponentially with momentum,
 the scale being fixed by the coupling constant $e^2$, which has dimension of
mass in our 3-Euclidean dimensions. We see that insisting on a perturbative
expansion led to the divergences. Defining the Green function through an
integral equation gave a result free of divergences. It falls off
much faster than the tree approximation which makes the pertubative calculation
diverge.

In this Note we argued that the divergences in QFT are due to presumptions
regarding the asymptotic behaviours of the Green function of the interacting
theory. Calculating Green functions using appropriate integral equation
evades the infinities. We used a
toy model to explicitly demonstrate our contentions. It is yet to be demonstrated
that this can be made into a systematic and a viable tool consistent with requirements
such as unitarity and causality in realistic theories. These issues will be addressed
elsewhere.

\end{document}